\documentclass[reprint,amsmath,amssymb,aps,prl,showpacs,floatfix]{revtex4-1}

\usepackage[dvips]{color,graphicx}
\DeclareGraphicsExtensions{.eps, .jpg, .png}
\usepackage{dcolumn}

\begin{document}

\title{Decomposition and terapascal phases of water ice}

\author{Chris J.\ Pickard} \email{c.pickard@ucl.ac.uk}
\affiliation{Department of Physics \& Astronomy, University College
  London, Gower Street, London WC1E~6BT, UK}

\author{Miguel Martinez-Canales}
\affiliation{Department of Physics \& Astronomy, University College
  London, Gower Street, London WC1E~6BT, UK}

\author{Richard J.\ Needs} 
\affiliation{Theory of Condensed Matter Group, Cavendish Laboratory,
  Cambridge CB3 0HE, UK}

\date{\today}

\begin{abstract}

  Computational searches for stable and metastable structures of water
  ice and other H:O compositions at TPa pressures have led us to
  predict that H$_2$O decomposes into H$_2$O$_2$ and a hydrogen-rich
  phase at pressures of a little over 5 TPa.
  The hydrogen-rich phase is stable over a wide range of hydrogen
  contents, and it might play a role in the erosion of the icy
  component of the cores of gas giants as H$_2$O comes into contact
  with hydrogen.
  Metallization of H$_2$O is predicted at a higher pressure of just
  over 6 TPa, and therefore H$_2$O does not have a thermodynamically
  stable low-temperature metallic form.
  We have also found a new and rich mineralogy of complicated water
  ice phases which are more stable in the pressure range 0.8--2 TPa
  than any predicted previously.
\end{abstract}

\pacs{64.70.K-, 71.15.Mb, 61.50.Ks, 62.50.-p}


\maketitle

Water ice under high pressures is an important component of gas giant
planets, and it has been speculated that it is present in the core of
Jupiter at pressures as high as 6.4 TPa \cite{Militzer_2008}.
The pressures at the centers of massive exoplanets can reach 10 TPa or
more \cite{exoplanets}, and establishing the properties of materials
at TPa pressures is a very difficult task.
Knowledge of the equation of state and whether the high-pressure
phases are insulating or metallic is particularly important.  

Hydrogen and oxygen are, respectively, the most abundant and the third
most abundant elements in the solar system \cite{Arnett_book}.  The
spatial distributions of elements within planets are understood to
some extent, but it is not in general known what chemical compounds
are stable at TPa pressures.
H$_2$O is a stable stoichiometry of the binary H:O system at low
temperatures and pressures, and the reaction
\begin{equation*}
{\rm H}_2{\rm O} \rightarrow {\rm H}_y{\rm O}_z + {\rm H}_{2-y}{\rm O}_{1-z}
\end{equation*}
is endothermic for all $y$ and $z$, $0 < y < 2, 0 < z < 1$.

TPa pressures are becoming more accessible experimentally, but it is
not currently possible to determine the stable stoichiometries of
materials at TPa pressures experimentally, or the crystalline
structures of any compounds formed.
We can, however, make progress using theoretical approaches.  The
capability to search for thermodynamically stable crystal structures
using density functional theory (DFT) methods has developed rapidly in
recent years.  Here we report searches for stable structures of
various H:O stoichiometries using DFT methods, which have allowed us
to investigate the stability of H$_2$O and of other stoichiometries at
TPa pressures.

Static-compression diamond-anvil-cell experiments have given us a
great deal of information about water ice up to pressures of 0.21 TPa
\cite{Hemley_1987,Goncharov_1996,Loubeyre_1999}, but this is far below
the highest pressures to which water is subjected within planets.
Shock wave experiments can reach much higher pressures, and sample
precompression \cite{Lee_2006,JeanlozCCELMBL07} and ramped compression
\cite{Davis_2005,HawreliakCEKLPRRSW07} techniques can reduce the
resulting temperatures to those more appropriate for planetary
science.  

The low-pressure and temperature phases of ice consist of packings of
hydrogen-bonded water molecules \cite{Salzmann_2009}.  Compression of
the high-pressure molecular ice VIII phase leads to a transition at
about 0.1 TPa to the ice X structure, in which the H atoms move to the
mid-points between neighbouring O atoms and the molecules lose their
separate identities \cite{Polian_ice_X_1984}.  Ice X is the
highest-pressure phase that has been observed experimentally, but DFT
studies have predicted further phase transitions at higher pressures
\cite{Benoit_1996,Militzer_2010,McMahon_2011_ice,Ji_2011,Wang_2011}.
The structural chemistry of some of these phases is discussed in Ref.\
\onlinecite{Hermann_2011}.

We have performed DFT calculations \cite{Supplemental} of structures
with various H:O stoichiometries using the \textsc{castep}
\cite{ClarkSPHPRP05} plane-wave code, the Perdew-Burke-Ernzerhof (PBE)
\cite{Perdew_1996_PBE} Generalized Gradient Approximation (GGA)
density functional, and ultrasoft pseudopotentials
\cite{Vanderbilt90}.  We searched for low enthalpy structures using
the \textit{ab initio} random structure searching (AIRSS) method
\cite{PickardN06_silane,Airss_review}.  This method has been applied
to many systems including H$_2$O at low pressures
\cite{Pickard_2007_water}, and hydrogen
\cite{pickard-h,McMahon_2011_hydrogen,Pickard_2012_hydrogen} and
oxygen \cite{Sun_2012_oxygen} at high pressures.  AIRSS involves
choosing starting structures and relaxing each of them to a minimum in
the enthalpy.  We have made extensive use of symmetry constraints in
our searches.  Starting structures were generated conforming to a
particular space group symmetry, although they were otherwise random,
and they were relaxed while maintaining the symmetry constraint.  We
performed searches for H$_2$O structures with up to 16 formula units
(fu), and searches for other stoichiometries were performed with up to
98 atoms \cite{Supplemental}.

The stability ranges of the most energetically favorable phases of
H$_2$O are given in Table \ref{table:pressure_stability}, including
those found in earlier DFT studies
\cite{Benoit_1996,Militzer_2010,McMahon_2011_ice,Ji_2011,Wang_2011}.
The AIRSS calculations produced three water ice structures of
symmetries $P3_121$ (or its chiral enantiomorph $P3_212$), $Pcca$ and
$C2$, with lower static-lattice enthalpies within the pressure range
0.78--2.36 TPa than those known previously, see Table
\ref{table:pressure_stability} and Ref.\ \onlinecite{Supplemental}.
We also found the $Pmc2_1$ structure reported by McMahon
\cite{McMahon_2011_ice} and the $I\bar{4}2d$ phase of Ref.\
\onlinecite{Wang_2011}, but they are metastable on our phase diagram.
At higher pressures we found the $P2_1$
\cite{McMahon_2011_ice,Wang_2011,Ji_2011}, $P2_1/c$ \cite{Ji_2011} and
$C2/m$ \cite{McMahon_2011_ice} structures of earlier studies.
Theoretical predictions of still higher pressure phases have been
reported but, as we show below, they are not stable \cite{Zhang_2012}.
We also performed calculations using the local density approximation
(LDA) \cite{Perdew_1981} for the $Pbca \leftrightarrow P3_121$ and
$P2_1/c \leftrightarrow C2/m$ transitions, which gave transition
pressures similar to those using the PBE functional
\cite{Supplemental}.  The LDA and PBE functionals have been tested
successfully in many high-pressure studies.  The pseudo-valence electronic
charge densities of materials become more uniform at very high
pressures, and the LDA and PBE functionals are particularly
appropriate under such conditions because they obey the uniform limit
and give an excellent description of the linear response of the
electron gas to an applied potential \cite{Perdew_1996_PBE}.
Additional discussion of these issues for carbon at TPa pressures is
presented in the Supplemental Material for Ref.\
\onlinecite{TPa_carbon}.

\begin{table}[ht!]
  \caption{\label{table:pressure_stability} 
    Space group symmetries, calculated stability ranges, and numbers 
    of fu per primitive unit cell for phases of H$_2$O.  Nuclear vibrational 
    motion is not included in this data.  The right-hand column gives the 
    source of the structure.}
\begin{ruledtabular}
\begin{tabular}{l|c|c|c}
Space group & Stability range (TPa) & No.\ fu & Source \\ \hline 
Ice X       &    --0.30             &   2     & Ref.\ \cite{Polian_ice_X_1984} \\
$Pbcm$      &    0.30--0.71         &    4    & Ref.\ \cite{Benoit_1996} \\ 
$Pbca$      &    0.71--0.78         &    8    & Ref.\ \cite{Militzer_2010} \\  
$P3_121$    &    0.78--2.01         &    12   & This work \\ 
$Pcca$      &    2.01--2.24         &    12   & This work \\ 
$C2$        &    2.24--2.36         &    12    & This work \\ 
$P2_1$      &    2.36--2.75         &    4    & Ref.\ \cite{McMahon_2011_ice,Wang_2011,Ji_2011}   \\
$P2_1/c$    &    2.75--6.06         &    8    & Ref.\ \cite{Ji_2011} \\ 
$C2/m$      &    6.06--             &    2   & Ref.\ \cite{McMahon_2011_ice}  \\ 
\end{tabular}
\end{ruledtabular}
\end{table}

\begin{figure}
  \centering
  \includegraphics[width=0.45\textwidth]{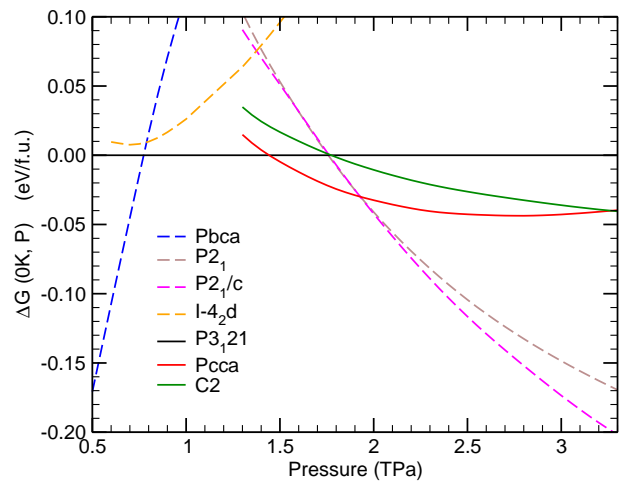}
  \caption{(Color online) Variation of the enthalpies with pressure of
    high-pressure water-ice structures. Previously-known structures
    are indicated by dashed lines, and phases that we
    have predicted are shown as solid lines.  ZP motion is included at the
    quasi-harmonic level.}
  \label{fig:enthalpies_with_zpe}
\end{figure}

\begin{figure}
  \centering
  \includegraphics[width=0.5\textwidth]{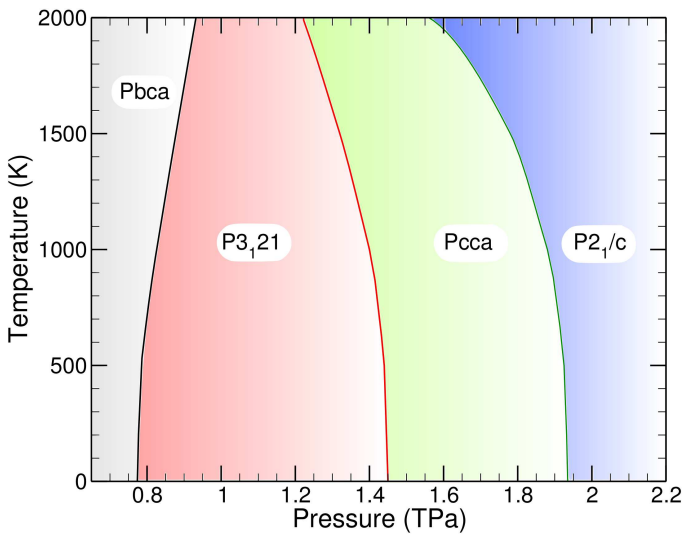}
  \caption{(Color online) Phase diagram of water ice including vibrational motion.}
  \label{fig:phase_diagram_with_vib}
\end{figure}

It is important to include the quantum nuclear zero-point (ZP) motion
when considering the energetics of systems containing light atoms.  We
therefore calculated the vibrational modes of the most stable phases
within the quasi-harmonic approximation and evaluated the
corresponding contributions to the free energies and pressures.
Technical details of the phonon calculations are given in the
Supplemental Material \cite{Supplemental}.  The enthalpy-pressure
relations of the relevant phases, including ZP motion, are shown in
Fig.\ \ref{fig:enthalpies_with_zpe}.  (The enthalpy reduction arising
from the $Pbcm$ distortion is too small to be visible on the scale of
Fig.\ \ref{fig:enthalpies_with_zpe}.) When ZP motion effects are
included the $C2$ and $P2_1$ phases no longer have regions of
thermodynamic stability, and our $P3_121$ and $Pcca$ phases become
stable in the ranges 0.77--1.44 TPa and 1.44--1.93 TPa, respectively.
Fig.\ \ref{fig:phase_diagram_with_vib} shows the computed phase
diagram including vibrational motion up to 2000 K.  Note that our
predicted pressure for the onset of stability of $P3_121$ H$_2$O of
0.77 TPa is not much higher than that of 0.7 TPa achieved in recent
shock compression of water \cite{Knudson_2012}, and that static
compression of rhenium up to 0.64 TPa was recently achieved in a
secondary anvil diamond cell \cite{Dubrovinsky_2012}, so that the
pressures considered here are likely to become accessible in the
future.

\begin{figure}
  \centering
  \includegraphics[width=0.275\textwidth]{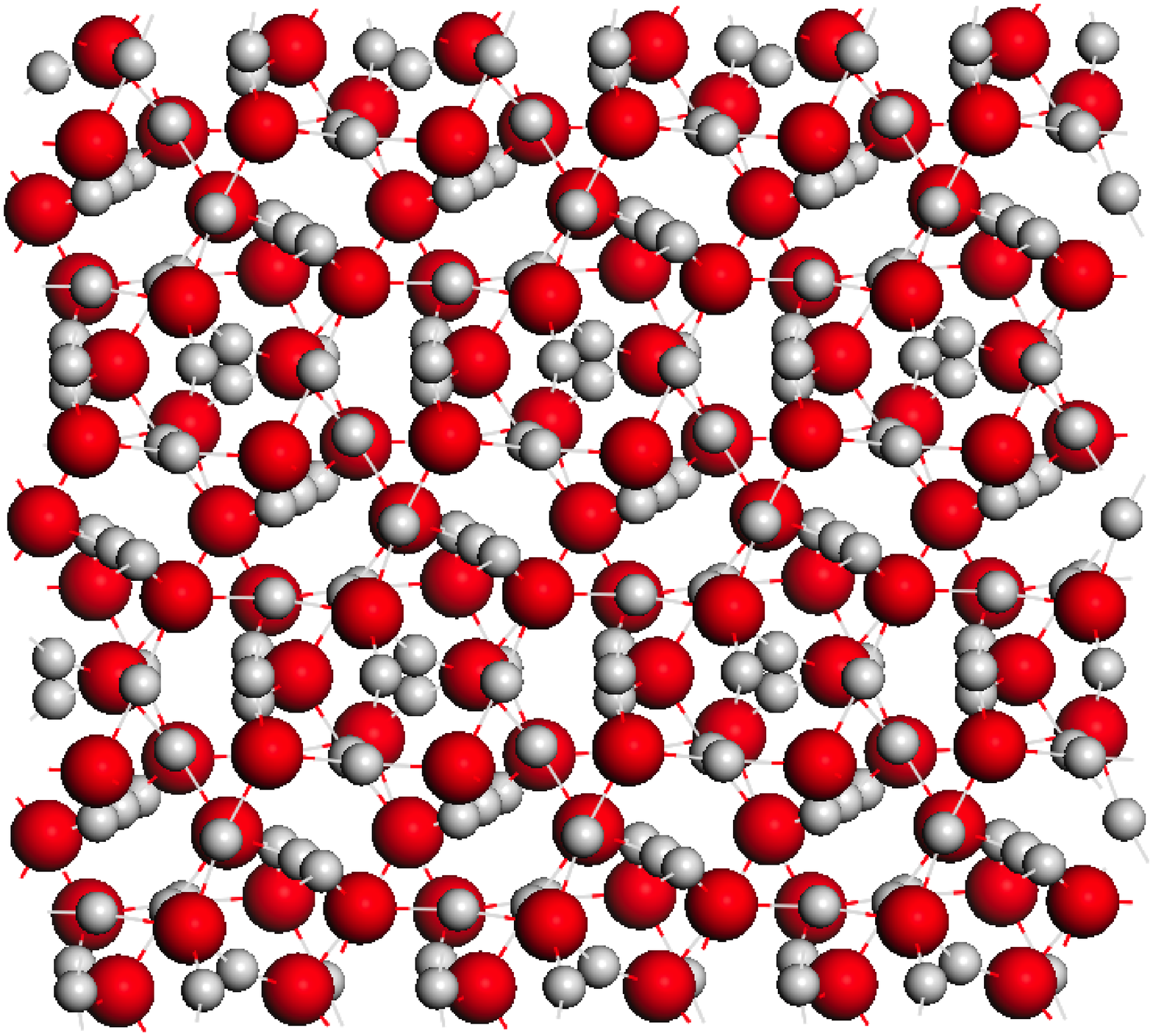}
  \includegraphics[width=0.25\textwidth]{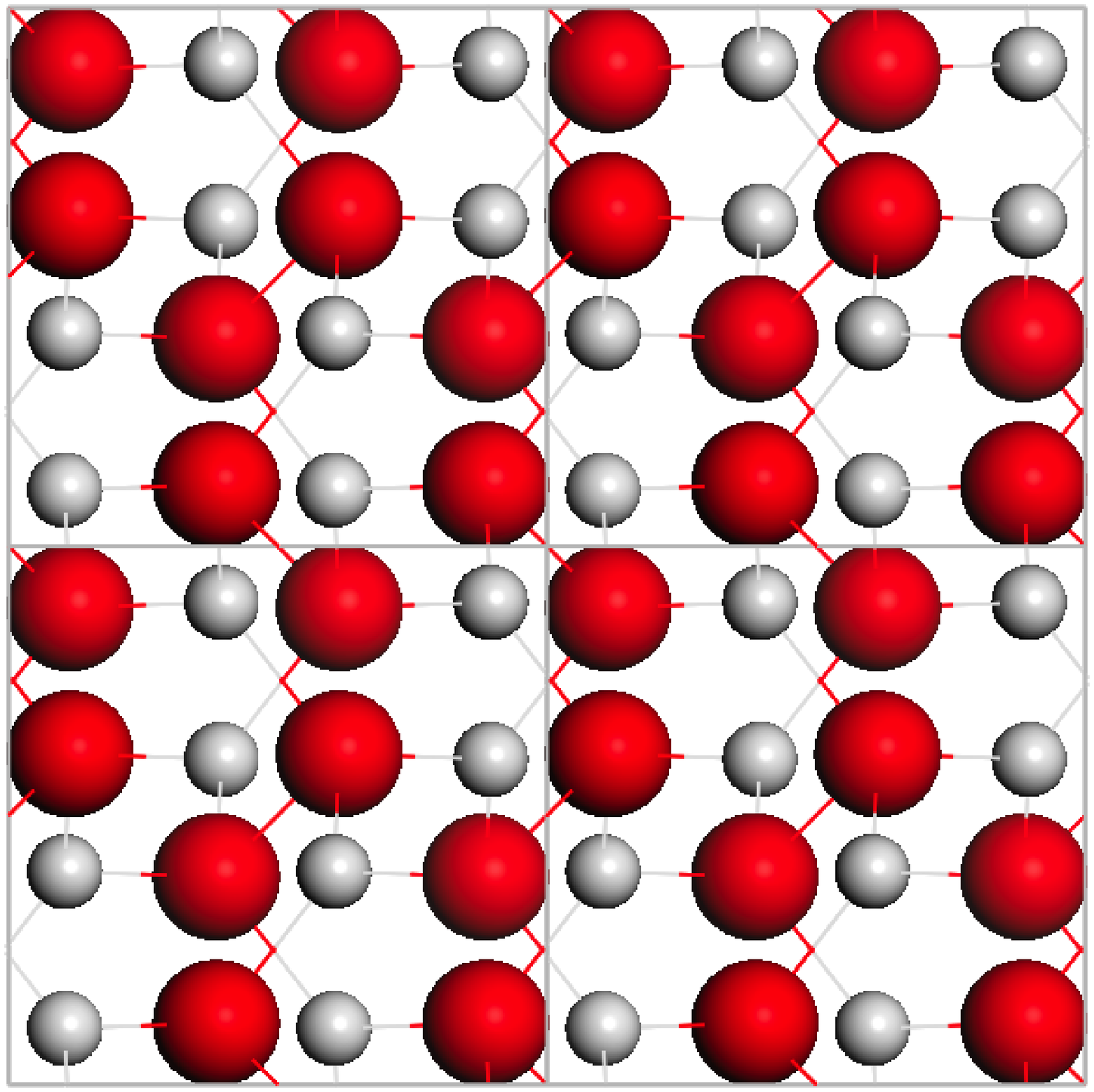}
  \caption{(Color online) (top) The $P3_121$ structure of water ice at 1
    TPa and (bottom) the $Pa\bar{3}$ structure of H$_2$O$_2$ at 6 TPa.
    The O atoms are shown in red and the H atoms in gray.}
  \label{fig:structures}
\end{figure}

The $Pbcm$, $Pbca$, $P2_1$, $P2_1/c$, $C2/m$, and $I\bar{4}2d$
structures have been described in previous work
\cite{Benoit_1996,Militzer_2010,McMahon_2011_ice,Ji_2011,Wang_2011}.
The O atoms of the $P3_121$ structure form hexagonal-close-packed
(hcp) layers, which are stacked not in the middle of the triangles of
the adjacent layers but mid-way along an edge, with a three-layer
repeat, see Fig.\ \ref{fig:structures}.
The density increases by about 2\% at the transition from $Pbca$ to
$P3_121$, which is reflected in the substantial reduction in the
gradient of the enthalpy-pressure curve at the transition apparent in
Fig.\ \ref{fig:enthalpies_with_zpe}.
The O lattice of the layered $Pcca$ structure exhibits a quartz-like
``bow-tie'' motif \cite{Supplemental}, and it is not particularly
similar to any of the standard close packed structures.  The primitive
cells of the $P3_121$ and $Pcca$ structures are large, containing 12
fu, and they appear to be of previously-unknown structure types.
The $C2$ structure also exhibits the bow-tie motif.
Details of the $P3_121$, $Pcca$ and $C2$ structures are reported in
the Supplemental Material \cite{Supplemental}.

H$_2$O and H$_2$ readily form hydrogen clathrate compounds at low
pressures \cite{Mao_2002}.  Much denser structures are favored at high
pressures, which can lead to changes in bonding and/or decomposition
into compounds of other stoichiometries, and a recent study has shown
that H:O compounds other than H$_2$O may be stable at TPa pressures
\cite{Zhang_2012}.  For example, we have identified a high-symmetry,
well-packed, very stable and insulating structure of hydrogen peroxide
(H$_2$O$_2$) of space group $Pa\bar{3}$, see Fig.\
\ref{fig:structures}, which contains O-O bonds and 3-fold coordinated
H atoms.  We have found that this phase plays an important role in
determining the stability of H$_2$O at high pressures.

We have searched over various H:O stoichiometries to investigate the
stability of H$_2$O to decomposition at TPa pressures.  We found an
instability of H$_2$O at pressures a little above 5 TPa to a
decomposition of the form
\begin{equation} 
  \label{eq:decomposition}
  {\rm H}_2{\rm O} \rightarrow \frac{\delta}{1+\delta}
  \frac{1}{2} \, {\rm H}_2{\rm O_2} + \frac{1}{1+\delta} \, {\rm
    H}_{2+\delta}{\rm O},
\end{equation}
with $\delta \geq 1/8$.
The right hand side of this reaction equation is simply H$_2$O when
$\delta=0$, but for $\delta>0$ it corresponds to the formation of
H$_2$O$_2$ and a hydrogen rich H$_{2+\delta}$O compound.

\begin{figure}
  \centering
  \includegraphics[width=0.45\textwidth]{Fig4-ConvexHull.eps}
  \includegraphics[width=0.5\textwidth]{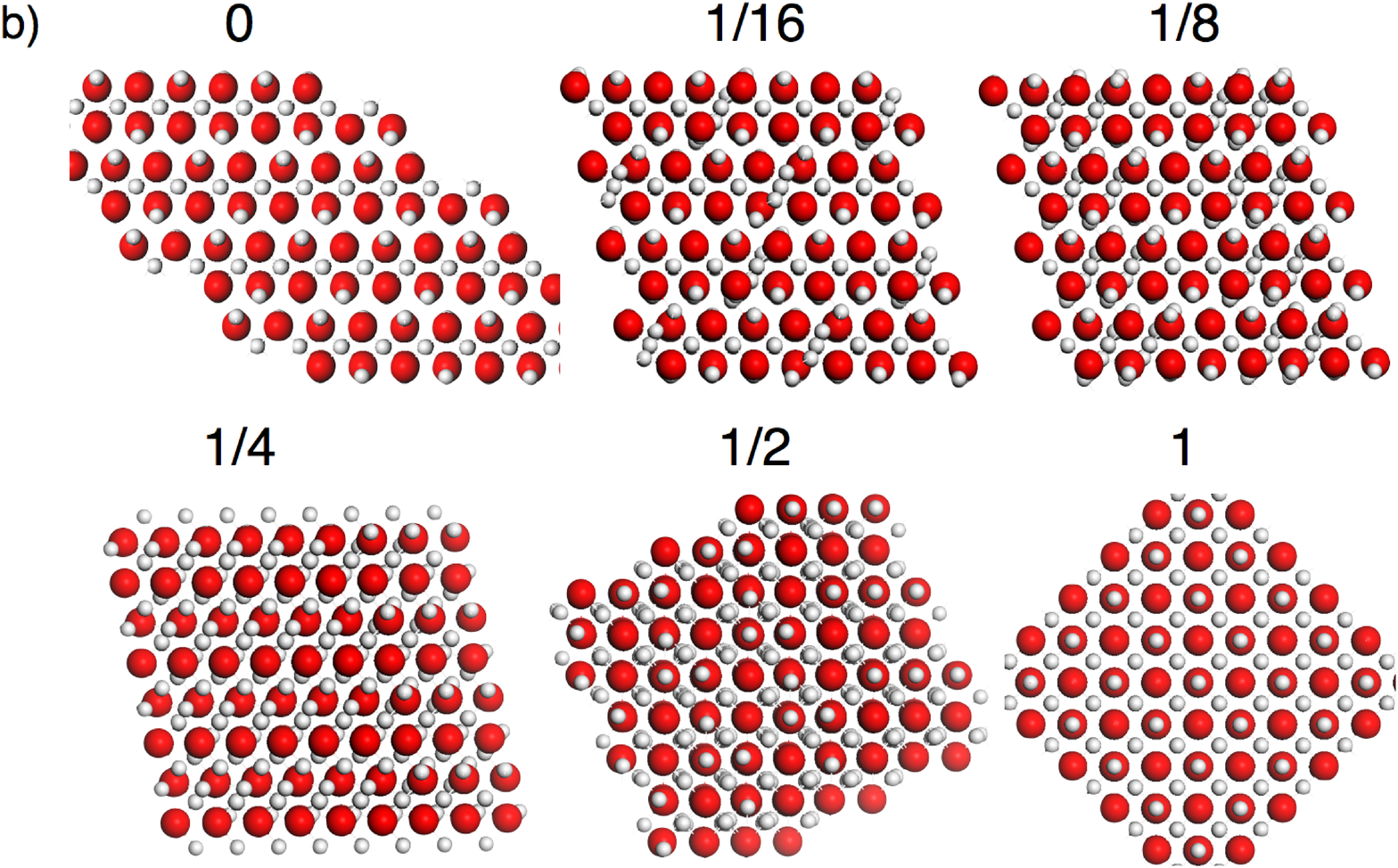}
  \caption{(Color online) (a) Convex hull diagram and enthalpies of
    ($\frac{1}{2}$H$_{2}$O$_{2}$)$_{1-x}$H$_x$ structures.  The red
    and black dotted lines show the convex hulls at 4 and 6 TPa,
    respectively.  The solid lines show enthalpies of stable
    structures at 4 TPa (red) and 6 TPa (black).  At 4 TPa the
    $P2_1/c$ phase (lower red dot at $x=0.5$) is more stable than
    $C2/m$ (upper red dot at $x=0.5$).  At 6 TPa the $P2_1/c$ and
    $C2/m$ phases are almost degenerate (black dot at $x=0.5$).  Six
    values of $\delta$ are labelled on the enthalpy curves, and the
    enthalpies of the corresponding structures for $\delta \simeq$ 1/8
    to $\delta \simeq$ 5/12 lie on the convex hull at 4 and 6 TPa.
    (b) Six H$_{2+\delta}$O structures at 6 TPa with values of
    $\delta$ from 0 to 1/2.  These structures have very similar
    arrangements of O atoms (red) for $\delta = 0$ to 1/4, but between
    $\delta = 1/4$ and $1/2$ the O packing changes to
    bcc.}
  \label{fig:c2/m-like_structures}
\end{figure}

This instability is illustrated in more detail in the convex hull
diagram of Fig.\ \ref{fig:c2/m-like_structures}(a).  At high pressures
the $Pa\bar{3}$ H$_2$O$_2$ structure is on the convex hull at $x=0$.
At 4 TPa, which is below the instability to decomposition, the
$P2_1/c$ phase of water ice is on the convex hull and is stable.  At 6
TPa the $P2_1/c$ and $C2/m$ phases are almost degenerate and the
convex hull passes below them, and all H$_2$O structures are unstable
to decomposition.  This decomposition is shown in Eq.\
\ref{eq:decomposition}, and it leads to the formation of H$_2$O$_2$ +
H$_{2+1/8}$O, so that $\delta=1/8$.  Note, however, that the
enthalpies of hydrogen rich structures from $\delta=1/8$ up to about
$\delta=5/12$ are on the convex hull and they are therefore also
stable at 6 TPa.  Some of the most stable structures that we have
found with fractional values of $\delta$ have large unit cells
containing up to 98 atoms.  It is likely that a quasi-continuum of
structures with different values of $\delta$ are stable at 6 TPa.  We
have also calculated the enthalpies of the structures at 5 TPa, and
find that this pressure is close to, but just below the lowest
pressure at which H$_2$O decomposes.  We therefore conclude that, at
low temperatures, H$_2$O becomes unstable to decomposition at
pressures of a little above 5 TPa.

From $\delta = 0$ up to somewhere between $\delta = 1/4$ and $1/2$ the
hydrogen-rich H$_{2+\delta}$O structures resemble the $C2/m$ phase,
but with layers of H atoms inserted, see Fig.\
\ref{fig:c2/m-like_structures}(b). We also found hydrogen deficient
$C2/m$ variants, but they are not stable under the conditions studied
here.  The appearance of structures with $\delta \neq 0$ can be
understood as a topotactic transition \cite{Shannon_topotaxy} to
structures in which the O lattice is maintained, while the $C2/m$-like
structures differ in the amount of hydrogen incorporated.  These
phases might be described as interstitial solid solutions.  

The H$_{2+\delta}$O phases are weakly metallic
\cite{Supplemental}. The relative enthalpies of the H$_{2+\delta}$O
phases correlate with the density of electronic states at the Fermi
energy (eDos($E_F$)).  The eDos($E_F$) takes its minimum value at
$\delta = 1/4$, which corresponds to the minimum in the convex hull of
Fig.\ \ref{fig:c2/m-like_structures}, see also Ref.\
\onlinecite{Supplemental}.  This suggests that the relative stability
of the H$_{2+\delta}$O phases is connected with the Fermi
surface/Bragg plane mechanism \cite{Jones_1934}, in which the energy
is reduced by the formation of a pseudo-gap in the eDos around $E_F$
arising from a symmetry breaking mechanism, which in this case
involves the insertion of H atoms.  It is possible that the true
ground states of the H$_{2+\delta}$O phases could be incommensurate
charge-density-wave metals with wavelengths related to the Fermi
surface.

If H$_2$O occurs under conditions of excess hydrogen, for example, at
the core/mantle boundary in a gas giant planet, the metallic
hydrogen-rich $C2/m$-like and bcc-like phases that we have found could
act as ``hydrogen sponges'', soaking up hydrogen from the mantle.  We
suggest that such a mechanism might play a role in erosion of the ice
component in the core of gas giants as H$_2$O comes into contact with
H$_2$.  Entropic effects also favor dissolution of H:O compounds in
hydrogen at high temperatures \cite{Wilson_2012}.

In summary, we have found a new and rich mineralogy of complicated
water ice phases of previously-unknown structure types, which lead to
a revision of the predicted phase diagram of H$_2$O within the
pressure range of about 0.8--2 TPa and above 5 TPa.
We predict that H$_2$O decomposes into H$_2$O$_2$ and hydrogen rich
phases based on the layered $C2/m$ structure of H$_2$O and on phases
with a bcc O lattice at pressures a little above 5 TPa.  This suggests
that H$_2$O is not a stable compound at the highest pressures at which
it has been suggested to occur within Jupiter.
We suggest that the $C2/m$ structure might play a role in the erosion
of icy cores of gas giant planets.
The insulator/metal transition is predicted to occur at the transition
from $P2_1/c$ to $C2/m$, but H$_2$O is unstable to decomposition at
this pressure, and therefore it does not have a thermodynamically
stable low-temperature metallic form.
Our study supports previous suggestions that icy planetary cores could
be strongly eroded by contact with a hydrogen-rich mantle
\cite{Stevenson_1982,Guillot_1999,Wilson_2012}.

\begin{acknowledgments}

  We acknowledge financial support from the Engineering and Physical
  Sciences Research Council (EPSRC) of the United Kingdom, and the use
  of the UCL Legion High Performance Computing Facility, and
  associated support services.  We thank Sian Dutton for helpful
  discussions.

\end{acknowledgments}

\end{document}